\documentclass[aip,jcp,superscriptaddress,floatfix,reprint]{revtex4-1}
\usepackage[utf8]{inputenc}
\usepackage{graphicx}
\usepackage{epsfig}
\usepackage{amsmath}
\usepackage{amssymb}
\usepackage{dcolumn}
\usepackage{braket}
\usepackage{multirow}
\usepackage[acronym]{glossaries} %
\usepackage[version=3]{mhchem}
\usepackage{threeparttable}
\usepackage{color}

\newcolumntype{d}[1]{D{.}{.}{#1} }
\newcommand{\la}{\langle}
\newcommand{\ra}{\rangle}
\newcommand{\bea}{\begin{eqnarray}}
\newcommand{\eea}{\end{eqnarray}}

\newcommand{\ch}{\hat H_e}

\newcommand{\chp}{\hat {H}_e^{(p)}}
\newcommand{\eq}[1]{Eq.~(\ref{#1})}

\newcommand{\rr}{\mathbf{r}}
\newcommand{\RR}{\mathbf{R}}
\newcommand*{\id}{{\normalfont\hbox{1\kern-0.15em \vrule width .8pt depth-.5pt}}}
\newcommand{\sinv}{(S^{-1})}
\newcommand{\thop}{\hat{\theta}_p}

\newcommand{\tilp}{\tilde{\theta}_p}

\newcommand{\YM}{\mathbf{Y}}

\newcommand{\XM}{\mathbf{X}}

\newcommand{\AM}{\mathbf{A}}
\newcommand{\BM}{\mathbf{B}}

\newcommand{\EPS}[1]{\epsilon_{#1}}

\newcommand{\FXC}{f_{\rm xc}^{\rm(DFT)}}
\newcommand{\EXC}{E_{\rm xc}^{\rm(DFT)}}
\newcommand{\EXCp}{E_{\rm xc}^{\rm(DFT,p)}}

\newcommand{\w}{\omega}

\newacronym{DFT}{DFT}{{D}ensity {F}unctional {T}heory}
\newacronym{KS-DFT}{KS-DFT}{{K}ohn--{S}ham density functional theory}
\newacronym{HF}{HF}{{H}artree-{F}ock}
\newacronym{KS}{KS}{{K}ohn--{S}ham}
\newacronym{AO}{AO}{atomic orbital}
\newacronym{MO}{MO}{molecular orbital}
\newacronym{RI}{RI}{resolution of identity}
\newacronym{TD-DFT}{TD-DFT}{time-dependent density functional theory}
\newacronym{EET}{EET}{electronic energy transfer}
\newacronym{LRC}{LRC}{long-range corrected}
\newacronym{LR-HF}{LR-HF}{long-range exact exchange}
\newacronym{LOPM}{LOPM}{localized operator partitioning method}
\newacronym{CT}{CT}{charge-transfer}
\newacronym{ET}{ET}{electron transfer}
\newacronym{DD}{DD}{difference density}

\begin{document}
\title{Localized operator partitioning method for electronic excitation energies in the time-dependent 
density functional formalism}
\author{Jayashree Nagesh}
\affiliation{Chemical Physics Theory Group, Department of Chemistry, University of Toronto, Toronto, ON M5S 3H6, Canada}
\author{Michael J. Frisch}
\affiliation{Gaussian, Inc., 340 Quinnipiac Street, Building 40, Wallingford, Connecticut 06492, U.S.A}
\author{Paul Brumer}
\affiliation{Chemical Physics Theory Group, Department of Chemistry, University of Toronto, Toronto, ON M5S 3H6, Canada}
\author{Artur F. Izmaylov}
\affiliation{Chemical Physics Theory Group, Department of Chemistry, University of Toronto, Toronto, ON M5S 3H6, Canada}
\affiliation{Department of Physical and Environmental Sciences, University of Toronto Scarborough, Toronto, ON M1C 1A4, Canada}

\date{\today}
\begin{abstract}
We extend the localized operator partitioning method  (LOPM) 
[J. Nagesh, A.F. Izmaylov, and P. Brumer, J. Chem. Phys. \textbf{142}, 084114 (2015)] 
to the time-dependent density functional theory (TD-DFT) framework 
to partition molecular electronic energies of excited states in a rigorous manner.
A molecular fragment is defined as a collection of atoms using 
Stratman-Scuseria-Frisch atomic partitioning. 
A numerically efficient scheme for evaluating the fragment excitation energy is 
derived employing a resolution of the identity to preserve standard one- and two-electron
integrals in the final expressions.
The utility of this partitioning approach is demonstrated by examining 
several excited states of two bichromophoric compounds: 
$9-$(($1-$naphthyl)$-$methyl)$-$anthracene and 
$4-$(($2-$naphthyl)$-$methyl)$-$benzaldehyde. 
The LOPM is found to provide nontrivial insights 
into the nature of electronic energy localization that 
are not accessible using simple density difference analysis.
\end{abstract}
\maketitle

\section{Introduction}
Understanding and controlling 
\gls{EET} is at the heart of effective utilization of solar energy,\cite{maykuhneet,*grossmanrsc2016} and
and efficient light harvesting in bio-molecular processes\cite{pach_pb2012}.
Therefore obtaining insights in \gls{EET} mechanisms through first principle modeling is
of paramount importance. A first-principles study of \gls{EET} is challenging because both electronic and nuclear 
degrees of freedom are usually involved (See for example Ref.~\onlinecite{subotnik2014,*bittner2014,*Barone:jctc2016}). 
Moreover, a quantitative investigation of \gls{EET} requires 
a computational tool that allows one to monitor how much electronic 
energy is located on a molecular fragment. 

When molecular fragments are well separated in space, various versions of the 
F\"orster theory can be successfully used for the \gls{EET} investigation.\cite{forster1959,Krueger:1998gq,Menucci:pccp2011} 
In contrast, monitoring electronic energy of a fragment becomes particularly challenging 
for flexible polymeric systems where electronic wave-function components of 
chromophoric fragments can overlap and thus interact relatively strongly (e.g., via $\pi-\pi$ stacking). 
Such overlaps make traditional approaches based on the F\"orster\cite{forster1959} 
and Dexter\cite{dexter} theories inadequate because of perturbative 
treatment of the inter-chromophore interactions and lack of rigorous spatial definition of the donor and acceptor. 
To avoid these deficiencies one can use techniques 
where multi-electronic state dynamics is obtained in the diabatic representation with a construction 
of diabatic states maximizing excitation localization but spanning the whole system.\cite{gmh1,*gmh2,*gmh_err,*fcd,*fed2,*fed3,*subotnik2010} 
However, such diabatic techniques do not quantify the amount of 
electronic energy located on a particular fragment of the molecule.\cite{izmaylovscience}

Recently, the \gls{LOPM} has been developed in order to 
address the problem of electronic energy partitioning independent of the degree of 
separation or strength of interaction between chromophoric fragments. The \gls{LOPM} is immune 
to the described problems since it formulates the partitioned energy electronic Hamiltonian that 
provides the corresponding local electronic energy as an
expectation value $E_p(t) = \langle\Psi(t)|H_p|\Psi(t)\rangle$ 
using the total system wave-function $|\Psi(t)\rangle$. That is, the \gls{LOPM} philosophy
is based on the quantum requirement that an operator, here $H_p$, 
corresponds to every measurable.  Note that both fully quantum 
electron-nuclear wave-function or electron only wave-function from mixed quantum-classical 
approaches can be used in the $E_p(t)$ expression. Using any approach to define the spatial volume around a nucleus as an atom\cite{pendas2006} and 
grouping such atoms into fragments\cite{nakai2007, bochicchio2005}  the \gls{LOPM} provides non-perturbative energy partitioning with atomic resolution. 
This technique can be applied for investigating the stationary states as a first step toward a full dynamical 
description. Any electronic structure technique can be used in the \gls{LOPM}, but to provide 
a proof of concept illustration the simplest excited state generation method,
configuration interaction singles (CIS) have been used in Ref.~\onlinecite{nagesh2015}.
It is well-known however that CIS overestimates excitation energies compared to \gls{TD-DFT} methods when compared to experiment.
Progress in linear response \gls{KS} \gls{TD-DFT} over last couple of decades
has shown that it is feasible to solve for the ground and singly excited valence states of the full 
system\cite{Stratman:1998/JCP/8218,Scuseria:1999/JPC/4782} with a good balance 
between accuracy and computational cost. Thus in this paper 
we develop the \gls{LOPM} within the \gls{TD-DFT} formalism and obtain spatially partitioned ground and excited state
electronic energies, leading to first-principles-based approach into EET in realistic molecules.

The remainder of this paper is organized as follows. Section \ref{sec:loc_op_theory} overviews the theory behind the 
LOPM and describes its extension to the density functional formalism for ground and excited states.  
In Sec. \ref{sec:implementation} we describe details of implementation. Section \ref{sec:results} reports 
the application of the LOPM to two bichromophoric organic molecules where singlet-singlet and triplet-triplet EET processes
have been previously studied. Section \ref{sec:conclusions} concludes by providing a summary and outlook for the LOPM.

\section{Theory}\label{sec:loc_op_theory}
\subsection{Localized operator partitioning method}
We briefly review the \gls{LOPM} from Refs.~\onlinecite{nagesh2015} and \onlinecite{yaser2012} to establish the notation for various quantities. 
Our starting point is the electronic Hamiltonian obtained after the Born-Oppenheimer separation
\bea
{\ch} &=& \sum_m h(\mathbf{r}_m) + \sum_{m>n} \frac{1}{|\mathbf{r}_{m}-\mathbf{r}_n|} \notag\\
&+& \sum_{k >l} \frac{Z_{k}Z_{l}}{|\mathbf{R}_{k}-\mathbf{R}_{l}|},  \label{eqn:mol_h}
\eea
where 
\bea
h(\mathbf{r}_m) &=& -\frac{1}{2} \nabla_m^2 - \sum_{k} \frac{Z_{k}}{|\mathbf{R}_{k}-\mathbf{r}_m|}
\eea
is the one-electron part, $\mathbf{r}_m$ and $\mathbf{R}_{k}$ are 
electronic and nuclear coordinates, $\nabla_m^2$ is an electronic Laplacian, 
and $Z_{k}$'s are nuclear charges.\cite{aunote}
The nuclear-nuclear repulsion term in Eq.~(\ref{eqn:mol_h}) does not contribute to the 
electronic excitation energies that are of the main interest to \gls{EET} and therefore
will be neglected below. 

For stationary states of ${\ch}$, the partitioned electronic Hamiltonian $\chp$ 
of subsystem $p$ is defined as\cite{nagesh2015}
\bea 
\chp &=&  \sum_m \theta_p(\mathbf{r}_m)h(\rr_m)
+\frac{1}{2}\sum_{m\neq n} \frac{\theta_{p}(\mathbf{r}_m)}{|\mathbf{r}_{m}-\mathbf{r}_n|},
\eea
where
\bea
\theta_p(\mathbf{r}_m) &=& \left\{ \begin{array}{ll} 1 & \mbox{if } \mathbf{r}_m \in p \\ 0 & \mbox{otherwise} \end{array} \right.
\eea
is the one-electron projection operator for subsystem $p$.
For electronic eigenstates $\Psi_I$ of ${\ch}$, subsystem energies are given by 
$E_p^{(I)}=\la\Psi_I|\chp|\Psi_I\ra$. 
Similarly, we define average subsystem electron populations for each electronic state $I$ as
\bea\label{eq:popp}
{\cal N}_I^{(p)} = \la\Psi_I| \sum_m \theta_p(\mathbf{r}_m)|\Psi_I\ra.
\eea
Owing to the completeness relation for one-electron projection operators\cite{yaser2012}
\begin{equation}
\label{eq:locpop}
\sum_p \theta_p (\rr) = \mathbf{1}_\rr, 
\end{equation}
the subsystem properties $E_p^{(I)}$ and ${\cal N}_p^{(I)}$ are additive
and are summed to corresponding total unpartitioned values. 

\subsection{Partitioning in Kohn--Sham density functional theory}


The ground state \gls{KS-DFT} energy is given by
\bea
E[\rho] = \bra{\Phi_{\rm KS}}\hat H_e \ket{\Phi_{\rm KS}}_S + \EXC [\rho],
\eea 
where $\ket{\Phi_{\rm KS}}$ is the KS determinant, 
subscript $S$ refers to scaling and introducing range separation in the \gls{HF} exchange part,
\begin{align}
\bra{\Phi_{\rm KS}}\hat H_e \ket{\Phi_{\rm KS}}_S &= \int d\rr \left[ h(\rr)\rho(\rr,\rr')\right]_{\rr=\rr'} \notag \\
&+ \sum_{n>m} \frac{Z_n Z_m}{|\RR_m-\RR_n|}  + J[\rho] + K[\rho], \label{eq:lrc-gs}\\
J[\rho] =\frac{1}{2}&\int d\rr d\rr' \frac{\rho(\rr')\rho(\rr)}{|\rr'-\rr|}, \\
K[\rho] = -\sum_{m}^{3} \frac{\alpha_m}{4} &\int d\rr d\rr' \frac{|\rho(\rr,\rr')|^2}{|\rr-\rr'|} {\cal O}_{m}(\rr,\rr'),
\end{align}
where ${\cal O}_m(\rr,\rr')$ represents long-range [$\text{erf}(\gamma|\rr-\rr'|)$], short-range [$\text{erfc}(\gamma|\rr-\rr'|)$] or full-range ($\mathbf{1}_{\rr,\rr'}$) operators with appropriate scaling factor $\alpha_m$, 
%
$\rho(\rr,\rr')$ is the one-particle density matrix 
corresponding to $\ket{\Phi_{\rm KS}}$ and $\rho(\rr')$ is its diagonal part.
The pure DFT exchange-correlation part is 
\bea
\EXC [\rho] = \int d\rr [e_x(\rho(\rr);\{\alpha_m\},\gamma) + e_c(\rho(\rr))],
\eea
where $e_x$ and $e_c$ are the exchange-correlation energy densities, 
the former parametrically depends on the scaling constants $\alpha_i$ and $\gamma$. 
Partitioning the scaled Hamiltonian in \eq{eq:lrc-gs} is done by  
partitioning the electronic Hamiltonian and then scaling the \gls{HF} exchange component
\bea\notag
\bra{\Phi_{\rm KS}}\hat H_{e}^{(p)}\ket{\Phi_{\rm KS}}_S
&=&  \int d\rr \theta_p(\rr)\left[ h(\rr)\rho(\rr,\rr')\right]_{\rr=\rr'} \notag \\
&& + J^{(p)}[\rho] + K^{(p)}[\rho] \label{eqn:partks} \\ 
J^{(p)}[\rho] &=& \frac{1}{2} \int d\rr d\rr' \frac{\rho(\rr')\rho(\rr)\theta_p(\rr)}{|\rr'-\rr|} \\
\notag
K^{(p)}[\rho] &=& -\sum_{m}^{3} \frac{\alpha_m}{4} \int d\rr d\rr' \frac{|\rho(\rr,\rr')|^2}{|\rr-\rr'|} \\
&&\times{\cal O}_{m}(\rr,\rr') \theta_p(\rr).
\eea
The pure DFT exchange-correlation part is partitioned term-wise 
by using the additivity of integration and $\theta_p(\rr)$ completeness [\eq{eq:locpop}]
\begin{equation}
\EXCp [\rho] = \int d\rr \theta_p(\rr)[ e_x(\rho(\rr);\{\alpha_i\},\gamma)+ e_c(\rho(\rr))].
\end{equation}
Therefore the partitioned \gls{KS} ground state energy is 
\bea
E^{(p)}[\rho] = \bra{\Phi_{\rm KS}}\hat H_e^{(p)} \ket{\Phi_{\rm KS}}_S + \EXCp [\rho].
\eea 

\subsection{Partitioning in time-dependent density functional theory}\label{subsec:parttd}

Casida's equations provide excitation energies, $\omega_I$, in a linear response regime as a 
solution of the generalized 
eigenvalue problem\cite{casida1995}
\begin{eqnarray}\label{eq:casmol}
\left( \begin{array}{cc}
\AM & \BM \\
{\BM}^{*} & {\AM}^{*} \end{array}\right) 
\left(\begin{array}{c}
\XM \\
\YM \end{array}\right)_I=
\w_I
\left( \begin{array}{cc}
\mathbf{1} & \mathbf{0} \\
 \mathbf{0} & \mathbf{-1} \end{array}\right)
\left(\begin{array}{c}
\XM \\
\YM \end{array}\right)_I,
\end{eqnarray}
where $\XM$ and $\YM$ stand for excitation and de-excitation coefficient vectors respectively.
The matrix elements of $\AM$ and $\BM$ written in the \gls{KS} \gls{MO} basis
are
\begin{align}
A_{ia,jb} &= \delta_{ij}\delta_{ab}(\EPS{a}-\EPS{i})
+ \la ij|\FXC|ab\ra \notag \\
&+ \la ij|ab\ra - \sum_m  \la ia|jb\ra_m, \label{eq:amo} 
\end{align}
and
\begin{equation}
B_{ia,jb} = \la ij|\FXC| ba\ra + \la ij|ba\ra - \sum_m \la ij|ab\ra_m, \label{eq:bmo}
\end{equation}
where $\EPS{a}$ and $\EPS{i}$ are \gls{KS} \gls{MO} energies, $\FXC$ is the exchange-correlation kernel
\begin{equation}
\FXC(\rr,\rr') =  \frac{\delta^2 \EXC}{\delta \rho(\rr) \delta \rho(\rr')},
\end{equation}
and 
\begin{align}
\la rs|tu\ra &= \int d\rr d\rr'\phi_r(\rr)\phi_t(\rr)\frac{1}{|\rr-\rr'|}\phi_s(\rr')\phi_u(\rr'), \\
\la rs|tu\ra_{m} &= \alpha_m \int d\rr d\rr'\phi_r(\rr)\phi_t(\rr)\frac{{\cal O}_m(\rr,\rr')}{|\rr-\rr'|}\phi_s(\rr')\phi_u(\rr') 
\end{align}
are the Coulomb and scaled exchange integrals respectively in Dirac's notation.
Here we use labels $i,j,\ldots$ for occupied; $a,b,\ldots$ for unoccupied,
and $r,s,\ldots$ to indicate either type of orbitals. 

It is convenient to recast the excitation energy as
\bea
\omega_I = (\XM^{\dagger}~\YM^{\dagger})_I
\left( \begin{array}{cc}
\AM & \BM \\
{\BM}^{*} & {\AM}^{*} \end{array}\right) 
\left(\begin{array}{c}
\XM \\
\YM \end{array}\right)_I. \label{eq:excten}
\end{eqnarray}
The origin of this quadratic form is the second variation of 
energy with respect to the one-electron density
\begin{equation}
\omega_I  = E_I - E_0 =  \int d\rr \int d\rr' \frac{\delta^2 E[\rho]}{\delta \rho(\rr) \delta \rho(\rr')} \rho_I^{(1)}(\rr) \rho_I^{(1)}(\rr'), 
\end{equation}
where $\rho_I^{(1)}(\rr)$ is the $I$th first order density response to an external potential variation, 
e.g., exciting laser field.
Since the total energy contains orbital dependent part, $\bra{\Phi_{\rm KS}} H_{e} \ket{\Phi_{\rm KS}}_S$, and the 
exchange correlation part, $\EXC [\rho]$, their variations are usually done differently using variations with 
respect to the KS orbitals for the  $\bra{\Phi_{\rm KS}} H_{e} \ket{\Phi_{\rm KS}}_S$ part and variation with respect to the 
density for the $\EXC [\rho]$ part
\begin{align}
\omega_I  &= \sum_{ijab}\big\{U_{ia}V_{jb} \notag \\
&\times \int d\rr \int d\rr' \frac{\delta^2\bra{\Phi_{\rm KS}} H_{e} \ket{\Phi_{\rm KS}}_S}{\delta \phi_i(\rr) \delta \phi_j(\rr')} \phi_a(\rr) \phi_b(\rr') \big\} \notag \\ 
&+  \int d\rr \int d\rr' \frac{\delta^2 \EXC[\rho]}{\delta \rho(\rr) \delta \rho(\rr')} \rho_I^{(1)}(\rr) \rho_I^{(1)}(\rr'), \label{eq:wI}
\end{align}
where $\rho_I^{(1)}(\rr) = \sum_{ia} X_{ia} \phi_i^{*}(\rr)\phi_a(\rr) + Y_{ia} \phi_a^{*}(\rr)\phi_i(\rr)$,
and $U_{ia}V_{jb}$ are four possible products $X_{ia}^*X_{jb}$, $Y_{ia}^*X_{jb}$, $X_{ia}^*Y_{jb}$, and $Y_{ia}^*Y_{jb}$  depending 
on which part of $\delta^2\bra{\Phi_{\rm KS}} H_{e} \ket{\Phi_{\rm KS}}_S / \delta \phi_i(\rr) \delta \phi_j(\rr')$, bra and/or ket, the orbital variation is taking place.
Equation (\ref{eq:wI}) maps solving Casida's equation to finding 
normal modes in density variations with the second variation of energy with respect to the density 
as the electronic energy Hessian.
 
Our partitioning approach can be straightforwardly generalized to \eq{eq:wI} by 
switching the order between the partitioning operation and the second variation
 \begin{align}
\w_{I}^{(p)} &= \int d\rr d\rr' \mathcal{P}\left\{ \frac{\delta^2 E}{\delta \rho(\rr) \delta \rho(\rr')} \right\}  \rho_I^{(1)}(\rr) \rho_I^{(1)}(\rr') \\
&= \int d\rr d\rr' \left\{\frac{\delta^2 \mathcal{P}\left\{E \right\}}{\delta \rho(\rr) \delta \rho(\rr')} \right\}  \rho_I^{(1)}(\rr) \rho_I^{(1)}(\rr').
\end{align}
$\mathcal{P}\left\{E \right\}$'s variation is done in the same way as for the total energy expression.  
This leads to the following expression in terms of \gls{KS} orbitals
 \begin{align}
\w_{I}^{(p)} &= (\XM^{\dagger}~\YM^{\dagger})_I
\left( \begin{array}{cc}
\AM^{(p)} & \BM^{(p)} \\
{\BM^{(p)}}^{*} & {\AM^{(p)}}^{*} \end{array}\right) 
\left(\begin{array}{c}
\XM \\
\YM \end{array}\right)_I,
\end{align}
where
\begin{align}
A_{ia,jb}^{(p)} &= \delta_{ij}F_{ab}^{(p)} - \delta_{ab}F_{ij}^{(p)} \notag \\
&+ \bra{ij}\FXC\ket{ab}^{(p)} +\la ij| ab\ra^{(p)} - \sum_m \la ia|jb\ra_m^{(p)} \\ 
B_{ia,jb}^{(p)} &= \bra{ij}\FXC\ket{ba}^{(p)} +\la ij|ba\ra^{(p)}- \sum_m \la ij|ab\ra_m^{(p)}. 
\end{align}
Here, $F_{rs}^{(p)}$ stands for the partitioned \gls{KS} Fock matrix,
\begin{align}
F_{st}^{(p)} &= h_{st}^{(p)} + \sum_{i}\big\{ 2\la is|it\ra^{(p)} - \sum_m \la ii|st\ra^{(p)}_m \big\}, \\ \label{h_int}
h_{st}^{(p)} &= \int d\rr \theta_p(\rr) \phi_s^*(\rr) h(\rr)\phi_t(\rr), \\
\notag
\end{align}
and
\begin{align}
\la rs|\FXC|tu\ra^{(p)} &=  \int d\rr d\rr' \theta_p(\rr) \frac{\delta^2 \EXC}{\delta \rho(\rr) \delta \rho(\rr')} \notag \\
& \times  \phi_r(\rr)^* \phi_t(\rr)\phi_s(\rr')^* \phi_u(\rr'), \label{fxc_int}\\  \notag
\la rs|tu\ra^{(p)}_{m} &= \alpha_m \int d\rr d\rr' \theta_p(\rr) \frac{{\cal O}_m(\rr,\rr')}{|\rr-\rr'|} \\ \label{LR_int}
&\times \phi_r(\rr)^* \phi_t(\rr)\phi_s(\rr')^* \phi_u(\rr'),\\ \notag
\la rs|tu\ra^{(p)} &= \int d\rr d\rr' \theta_p(\rr) \frac{1}{|\rr-\rr'|} \\ \label{2e_int} 
&\times \phi_r(\rr)^* \phi_t(\rr)\phi_s(\rr')^* \phi_u(\rr').
\end{align}
We note that although the full Fock matrix is diagonal, its partitioned counterpart has non-zero off-diagonal elements.

\section{Implementation}\label{sec:implementation}
\subsection{Resolution of identity}
For efficient implementation of the partitioned energies within the \gls{TD-DFT} formalism
we use the localized \gls{AO} Gaussian basis set ($\phi_{\mu},\phi_{\nu}\ldots$) 
to evaluate the fragment ground and excited state energies. 
 This allows us to employ numerous screening techniques in 
generating  \gls{AO} one- and two-electron integrals that are contracted on-the-fly 
with corresponding densities.\cite{Scuseria:1999/JPC/4782} 
The direct scheme leads to AO counterparts of integrals in Eqs.~\eqref{h_int}-\eqref{2e_int}.
Partitioning the exchange-correlation part in $\FXC$ integrals [\eq{fxc_int}] is done by eliminating the quadrature points 
located outside of the subsystem region. Straightforward partitioning of the nuclear-electron attraction 
and electron-electron repulsion integrals would involve modification of the Boys integrals 
to accommodate the altered shape of the partitioned \glspl{AO} and thus would create computational difficulties.
To circumvent this problem we replace the partitioning operator $\thop$ 
by its projected form employing the \gls{RI} technique\cite{nagesh2015} 
with the projection operator
\begin{equation}
\hat{\id} = \sum_{\mu\nu} |\mu\rangle \sinv_{\nu\mu} \langle \nu|, \label{eqn:ri1e}
\end{equation}
where $\sinv_{\nu\mu}$ are matrix elements of the inverse of the \gls{AO} overlap matrix 
$S_{\mu\nu} = \bra{\mu}\nu\rangle$ and $|\mu\rangle, |\nu\rangle$ are \gls{AO} basis functions.
Then the projected form of $\thop$ is
\begin{align}\label{eq:tilp}
\tilp &= \hat{\id} ~\thop \hat{\id} \\
&= \sum_{\mu\nu} |\mu\ra L_{\mu\nu}^{(p)} \la \nu|,
\end{align}
where
\begin{align}
L_{\mu\nu}^{(p)} &= \sum_{\mu_1\nu_1} \sinv_{\mu \mu_1} S^{(p)}_{\mu_1 \nu_1} \sinv_{\nu_1 \nu}, \\
S^{(p)}_{\mu_1 \nu_1} &= \la \mu_1 | \thop | \nu_1\ra \\
&= \int d\rr \phi_{\mu_1}(\rr)\theta_{p}(\rr)\phi_{\nu_1}(\rr)\label{eq:Sp}.
\end{align}
Thus, to obtain the projected partitioning $\tilp$, the partitioned overlap matrix $S^{(p)}_{\mu\nu}$ elements are evaluated as weighted sums
\begin{equation}
S_{\mu\nu}^{(p)} = \sum_{k\in\Omega_p} \sum_i w(\rr_i) p_k(\rr_i) \phi_{\mu}(\rr_i)\phi_{\nu}(\rr_i),
\end{equation}
where $\Omega_p$ denotes a group of atoms representing fragment $p$ in the molecule, with $p_k(\rr_i)$ 
being $k^{\text{th}}$ atom's spatial partition function in the  Stratman-Scuseria-Frisch atomic partitioning scheme,
and $w(\rr_i)$ is the normalized quadrature weight associated with grid point $\rr_i$ in a spherical quadrature scheme. \cite{Scuseria1996}

Using the projected partitioning we recast the partitioned one-electron operator $\hat h$ \gls{AO} matrix elements 
as a matrix product of standard one-electron integrals and overlap matrices
\begin{equation}\label{eq:hpt}
\la \mu |\hat h\tilp |\nu \ra =  \sum_{\lambda\sigma} \la \mu |\hat h|\lambda \ra S_{\lambda\sigma}^{-1}  S_{\sigma\nu}^{(p)}, 
\end{equation}
similar products are obtained for the two-electron \gls{AO} integrals
\begin{align}
\la\mu\lambda|\nu\sigma\ra^{(p)} &= \sum_{\mu_1\nu_1} S_{\mu\mu_1}^{(p)}\sinv_{\mu_1\nu_1}\la\nu_1\lambda|\nu\sigma\ra \label{eq:gpart} \\
\la\mu\lambda|\nu\sigma\ra^{(p)}_{m} &= \sum_{\mu_1\nu_1} S_{\mu\mu_1}^{(p)}\sinv_{\mu_1\nu_1}\la\nu_1\lambda|\nu\sigma\ra_{m}. \label{eq:gpart_LR} 
\end{align}

Although there is a difference in results of partitioning by $\tilp$ and $\thop$,  this is not an issue because 
we consider $\tilp$ as our primary partitioning operator. This projected partitioning is more convenient
in implementation and  gives exactly the same partitioned density matrices as a non-projected version.

Thus, the projecting partitioning operator $\tilde{\theta}_p$ [\eq{eq:tilp}] allows us to use 
standard integrals in both one- and two-electron contributions [Eqs.~(\ref{eq:hpt})-(\ref{eq:gpart_LR})],  
and thus to by-pass the problem of partitioning the Boys integrals.  

\subsection{Working equations for \gls{TD-DFT} partitioned excitation energies in the \gls{AO} representation}
Before introducing partitioning in working energy expressions 
we provide those for the unpartitioned ground and excited states in the \gls{AO} representation. 
The ground state \gls{KS-DFT} energy for the closed shell case is given by
\begin{align}
E_{KS}^{(0)} &= \sum_{\mu\nu}P_{\mu\nu}^{(0)} h_{\mu\nu}+\EXC\notag \\
&+ \sum_{\mu\nu\lambda\sigma}\big\{ 2P_{\mu\nu}^{(0)}P_{\lambda\sigma}^{(0)}\la\mu\lambda|\nu\sigma\ra -   P_{\mu\sigma}^{(0)}P_{\lambda\nu}^{(0)}\sum_m \la\mu\lambda|\nu\sigma\ra_m \big\} \notag \label{eq:e0ks-ao}\\
\end{align}
where $P_{\mu\nu}^{(0)} = \sum_{i} C_{\mu i} C_{\nu i}$ is the AO density matrix 
and $\{C_{\mu p}\}$ are the \gls{KS} \gls{MO} coefficients.

The excited state energies $\omega_I$ are evaluated using the symmetrized and 
anti-symmetrized \gls{AO} transition densities $\mathbf{T^{\pm}}$, 
\begin{equation}
T_{\mu\nu}^{\pm(I)} = \sum_{ia} (X_{ia}^{(I)}\pm Y_{ia}^{(I)}) (C_{\mu i} C_{\nu a}\pm C_{\mu a} C_{\nu a}).
\end{equation}
Assuming a closed shell singlet state, we transform Eq.~\eqref{eq:wI} 
into the \gls{AO} basis and group the one- and two-electron terms to obtain
\begin{align}
\omega_I &= E_{orb}^{(I)}+ E_{xc}^{(I)} + E_{2e}^{(I)},\label{eq:es-ao} \\
E_{orb}^{(I)} &= \sum_{\mu\nu} R_{\mu\nu}^{(I)} h_{\mu\nu} \notag \\
&+\sum_{\mu\nu\lambda\sigma} \big\{ J_{\mu\nu\lambda\sigma}^{(I)}\la\mu\lambda|\nu\sigma\ra+\sum_m  K_{\mu\nu\lambda\sigma}^{(I)}\la\mu\lambda|\nu\sigma\ra_m 
\big\},\\
E_{xc}^{(I)} &= \sum_{\mu\nu\lambda\sigma} D_{\mu\nu\lambda\sigma}^{(I)} \la\mu\lambda|\FXC|\nu\sigma\ra,  \\ 
E_{2e}^{(I)} &= \sum_{\mu\nu\lambda\sigma} \big\{D_{\mu\nu\lambda\sigma}^{(I)}\la\mu\lambda|\nu\sigma\ra + \frac{1}{2}\Gamma_{\mu\nu\lambda\sigma}^{(I)}\sum_m \la \mu\lambda|\nu\sigma\ra_m \big\},
\end{align}
and 
\begin{align}
R_{\mu\nu}^{(I)} &= \sum_{ijab} U_{ia}^{(I)} U_{jb}^{(I)} \{\delta_{ij}C_{\mu b} C_{\nu a} - \delta_{ab} C_{\mu i} C_{\nu j}\},\\
J_{\mu\nu\lambda\sigma}^{(I)} &= 2 P_{\mu\nu}^{(0)} R_{\lambda\sigma}^{(I)} + 2 R_{\mu\nu}^{(I)}P_{\lambda\sigma}^{(0)},\\ 
K_{\mu\nu\lambda\sigma}^{(I)} &= - P_{\mu\sigma}^{(0)} R_{\lambda\nu}^{(I)} -  R_{\mu\sigma}^{(I)}P_{\lambda\nu}^{(0)},\\
D_{\mu\nu\lambda\sigma}^{(I)} &= T_{\mu\nu}^{+(I)} T_{\sigma\lambda}^{+(I)} +   T_{\lambda\sigma}^{+(I)} T_{\mu\nu}^{+(I)},\\
\Gamma_{\mu\nu\lambda\sigma}^{(I)} &=  \big\{ T_{\mu\nu}^{+(I)}T_{\sigma\lambda}^{+(I)} + T_{\mu\lambda}^{+(I)} T_{\nu\sigma}^{+(I)} + T_{\lambda\sigma}^{+(I)} T_{\nu\mu}^{+(I)} \notag \\
&+ T_{\lambda\mu}^{+(I)}T_{\sigma\nu}^{+(I)} + T_{\mu\lambda}^{-(I)}T_{\nu\sigma}^{-(I)} - T_{\mu\sigma}^{-(I)}T_{\lambda\nu}^{-(I)} \notag \\
&+ T_{\lambda\mu}^{-(I)}T_{\sigma\nu}^{-(I)} - T_{\lambda\nu}^{-(I)}T_{\mu\sigma}^{-(I)}\big\}. 
\end{align}
Therefore all components of excitation energy [\eq{eq:es-ao}] can be efficiently generated by contracting 
2-index density-like quantities with the standard one- and two-electron integrals.\cite{Izmaylov:2006/JCP/104103,Scuseria:1999/JPC/4782} 

The ground state energy is partitioned as
\begin{align}
E_{KS}^{(0,p)} &= \sum_{\mu\nu}\tilde{P}_{\mu\nu}^{(0,p)} h_{\mu\nu}+\EXCp\notag \\
+ \sum_{\mu\nu\lambda\sigma}&\big\{ 2\tilde{P}_{\mu\nu}^{(0,p)}P_{\lambda\sigma}^{(0)}\la\mu\lambda|\nu\sigma\ra -  \tilde{P}_{\mu\sigma}^{(0,p)}P_{\lambda\nu}^{(0)}\sum_m\la\mu\lambda|\nu\sigma\ra_m \big\} \label{eq:e0kspart-ao} 
\end{align}
where the first and last two terms are obtained by contracting the matrix elements from  Eqs.~(\ref{eq:hpt}), (\ref{eq:gpart}) and (\ref{eq:gpart_LR})  with $\mathbf{P}^{(0)}$ as
\begin{align}
\sum_{\mu\nu} P_{\mu\nu}^{(0)} \la\mu|\tilp \hat{h}|\nu\ra &= \sum_{\mu\nu\lambda\sigma} P_{\mu\nu}^{(0)} S_{\mu\lambda}^{(p)} \sinv_{\lambda\sigma} h_{\sigma\nu}  \notag \\
&= \sum_{\nu\sigma} \tilde{P}_{\nu\sigma}^{(0,p)} h_{\sigma\nu}, \label{eqn:part1_gs}\\
\sum_{\mu\nu\lambda\sigma}2 P_{\mu\nu}^{(0)}P_{\lambda\sigma}^{(0)}\la\mu\lambda|\nu\sigma\ra^{(p)} &= 
 \sum_{\mu\nu\lambda\sigma}2P_{\mu\nu}^{(0)}P_{\lambda\sigma}^{(0)}  \notag \\
 &\times \sum_{\mu_1\nu_1} S_{\mu\mu_1}^{(p)}\sinv_{\mu_1\nu_1}\la\nu_1\lambda|\nu\sigma\ra \\
&= \sum_{\nu_1\lambda\nu\sigma} 2\tilde{P}_{\nu\nu_1}^{(0,p)} \tilde{P}_{\lambda\sigma}^{(0)} \la\nu_1\lambda|\nu\sigma\ra.
\end{align}

Following a similar procedure for the excitation energies, we obtain
\begin{align}
\omega_I^{(p)} &= E_{orb}^{(I,p)}+ E_{xc}^{(I,p)} + E_{2e}^{(I,p)},\label{eq:espart-ao} \\
E_{orb}^{(I,p)} &= \sum_{\mu\nu} \tilde{R}_{\mu\nu}^{(I,p)} h_{\mu\nu}  \notag \\
&+\sum_{\mu\nu\lambda\sigma} \big\{ \tilde{J}_{\mu\nu\lambda\sigma}^{(I,p)}\la\mu\lambda|\nu\sigma\ra +\tilde{K}_{\mu\nu\lambda\sigma}^{(I,p)}\sum_m \la\mu\lambda|\nu\sigma\ra_m\big\} \label{eqn:eorbp} \\
E_{xc}^{(I,p)} &= \sum_{\mu\nu\lambda\sigma} \tilde{D}_{\mu\nu\lambda\sigma}^{(I,p)} \la\mu\lambda|\FXC|\nu\sigma\ra,  \label{eqn:ehxcp} \\ 
E_{2e}^{(I,p)} &= \sum_{\mu\nu\lambda\sigma} \big\{ \tilde{D}_{\mu\nu\lambda\sigma}^{(I,p)} \la\mu\lambda|\nu\sigma\ra +\frac{1}{2}\tilde{\Gamma}_{\mu\nu\lambda\sigma}^{(I,p)}\sum_m\la\mu\lambda|\nu\sigma\ra_m \big\}, \label{eqn:e2ep}
\end{align}
where
\begin{align}
\tilde{R}_{\mu\nu}^{(I,p)} &= \sum_{\lambda\sigma} R_{\mu\lambda}^{(I)}S_{\lambda\sigma}^{(p)}\sinv_{\sigma\nu},\\
\tilde{J}_{\mu\nu\lambda\sigma}^{(I,p)} &= 2 \tilde{P}_{\mu\nu}^{(0,p)} R_{\lambda\sigma}^{(I)} + 2 \tilde{R}_{\mu\nu}^{(I,p)}P_{\lambda\sigma}^{(0)},\\ 
\tilde{K}_{\mu\nu\lambda\sigma}^{(I,p)} &= - \tilde{P}_{\mu\sigma}^{(0,p)} R_{\lambda\nu}^{(I)} - \tilde{R}_{\mu\sigma}^{(I,p)}P_{\lambda\nu}^{(0)},
\end{align}
and using symmetrized $\tilde{\mathbf{T}}^{+(I,p)}$ and anti-symmetrized $\tilde{\mathbf{T}}^{-(I,p)}$ partitioned transitioned densities 
\begin{align}
\tilde{T}_{\mu\nu}^{(I,p)} &= \sum_{\lambda\sigma} T_{\mu\lambda}^{(I)}S_{\lambda\sigma}^{(p)}\sinv_{\sigma\nu},\\
\tilde{T}_{\mu\nu}^{\pm (I,p)} &= \frac{1}{2}[\tilde{T}_{\mu\nu}^{(I,p)} \pm \tilde{T}_{\nu\mu}^{(I,p)}],
\end{align}
the two-electron densities in $E_{Hxc}^{(I,p)}$ and $E_{2e}^{(I,p)}$ are
\begin{align}
\tilde{D}_{\mu\nu\lambda\sigma}^{(I,p)} &= \tilde{T}_{\mu\nu}^{+(I,p)} T_{\sigma\lambda}^{+(I)} +   \tilde{T}_{\lambda\sigma}^{+(I,p)} T_{\mu\nu}^{+(I)},\\
\tilde{\Gamma}_{\mu\nu\lambda\sigma}^{(I,p)} &= \big\{ \tilde{T}_{\mu\nu}^{+(I,p)}T_{\sigma\lambda}^{+(I)} + \tilde{T}_{\mu\lambda}^{+(I,p)} T_{\nu\sigma}^{+(I)} + \tilde{T}_{\lambda\sigma}^{+(I,p)} T_{\nu\mu}^{+(I)} \notag \\
&+ \tilde{T}_{\lambda\mu}^{+(I,p)}T_{\sigma\nu}^{+(I)} + \tilde{T}_{\mu\lambda}^{-(I,p)}T_{\nu\sigma}^{-(I)} - \tilde{T}_{\mu\sigma}^{-(I,p)}T_{\lambda\nu}^{-(I)} \notag \\
&+ \tilde{T}_{\lambda\mu}^{-(I,p)}T_{\sigma\nu}^{-(I)} - \tilde{T}_{\lambda\nu}^{-(I,p)}T_{\mu\sigma}^{-(I)}\big\}. 
\end{align}
Thus, using the projected local electronic partitioning allows us to formulate all computationally intense partitioned quantities as a product of standard Gaussian integrals contracted with various densities. These partitioned energy equations have been implemented in the Gaussian suite of programs.\cite{gdv-i06} 

\section{Results and Discussion}\label{sec:results}

\begin{figure}
\caption{A1N and Closs-M molecules with fragment definitions used in this work. Atoms in blue correspond to the donor in an \gls{EET} process.}
\includegraphics[scale=0.6]{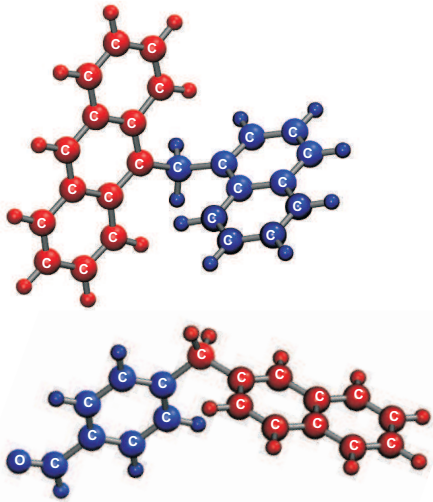}
\label{fig:bichrom}
\end{figure}

We apply the resulting formulae to partition energies of two bichromophores: $9-$(($1-$naphthyl)$-$methyl)$-$anthracene (A1N)\cite{levy2000} and $4-$(($2-$naphthyl)$-$methyl)$-$benzaldehyde (Closs-M)\cite{closs1988} (Fig.~\ref{fig:bichrom}), 
whose singlet-singlet and triplet-triplet \gls{EET} have been investigated experimentally. 
All TD-DFT calculations used 6-31G(d) basis set for energy calculations and the \gls{RI} expansion in the partioning
operator definition. The unrelaxed one-electron densities for excited states were used to analyze various properties 
of these states. A spherical quadrature scheme consisting of a pruned grid of 75 radial shells and 302 angular 
points per shell per atom\cite{lebedev76} was employed to evaluate numerical partitioned overlap integrals and DFT 
contributions. 

\begin{table*}
\caption{Comparison of \gls{TD-DFT} and TD-HF 
excitation energies, $\omega_I$ (eV), and oscillator strengths ($f_I$) 
in A1N using different functionals for 5 lowest excited states; 
the experimental vertical transition estimates are 3.34~eV (S$_0$-S$_1$) and 4.01~eV (S$_0$-S$_2$). 
The results for each functional are at the corresponding optimized geometry for the ground state.}\label{tab:func}
\begin{tabular}{cccccccccccccccc}
\toprule
State & &\multicolumn{2}{c}{LSDA}\cite{lsda1,*lsda2,*lsda3,*lsda4} & &\multicolumn{2}{c}{PBE1PBE}\cite{pbe1pbe1999} && \multicolumn{2}{c}{CAM-B3LYP}\cite{camb3lyp2004} && \multicolumn{2}{c}{$\omega$B97X-D}\cite{wb97xd2008} && \multicolumn{2}{c}{HF} \\
\hline
 & &$\omega_I$ & $f_I$  && $\omega_I$ & $f_I$ & & $\omega_I$ & $f_I$ && $\omega_I$ & $f_I$ && $\omega_I$ & $f_I$ \\
\hline
1 && 2.65 & 0.0039 && 3.26 & 0.1060&& 3.58 & 0.1411 && 3.59 & 0.1402&& 4.01 & 0.1988\\
 2& &2.85 & 0.0636 && 3.62 & 0.0007&& 4.10 & 0.0009&& 4.10 & 0.0011&& 4.63 & 0.0008\\
 3& &2.96 & 0.0024 && 3.88 & 0.0002&& 4.62 & 0.0065&& 4.66 & 0.0002&& 5.02 & 0.1039\\
 4& &3.50 & 0.0004 && 3.98 & 0.0011&& 4.70 & 0.0242&& 4.71 & 0.1217&& 5.29 & 0.0004\\
 5& &3.64 & 0.0006 && 4.45 & 0.1057&& 4.74 & 0.0929&& 4.81 & 0.0026&& 5.79 & 0.0044\\
\hline
\end{tabular}
\end{table*}

\textit{Total excitation energies of A1N:}
Table~\ref{tab:func} presents full system excited state energies using TD-HF, pure, 
hybrid, and \gls{LRC} functionals for A1N. 
The obtained results can be compared with available 
gas-phase experimental estimates obtained from fluorescence excitation spectra.\cite{levy2000} 
The original spectra are vibrationally resolved and to obtain estimates 
corresponding to vertical electronic transitions we used intensity-weighted 
sums $\omega_J^{\rm(exp)} = \sum_n I_n \Delta E_n$,\cite{davidson1998} 
where $\Delta E_n$ are the vibronic peak positions for the ground to $J$th excited electronic state transition, 
and $I_n$ are corresponding normalized intensities. 
Based on vertical excitation energies of bright transitions, 
PBE1PBE has the best agreement with experimental estimates. 
Nature of excited states was further analyzed using the atomic partitioning to 
obtain a fraction of electronic charge transferred from the anthracene upon the excitation, 
$\Delta\mathcal{N}_I^{(A)}$ [\eq{eq:popp}] in Table~\ref{tab:a1nlopm}.
This analysis shown that the dark states between two bright states in pure and hybrid 
functionals have a charge transfer (CT) character. This is consistent with previous studies revealing 
a problem of energy underestimation for CT states due to inaccuracies in a
treatment of electron-hole attraction in CT states. This problem is somewhat 
reduced in the long-range corrected functionals, CAM-B3LYP and $\omega$B97X-D,\cite{herbert2008,adamo2008,nitta2012} 
but their bright state excitations deviate from experiment by an amount more than that of PBE1PBE.

\begin{table}
\caption{Local \gls{TD-DFT} and TD-HF 
populations and excitation energies in A1N: $\Delta {\cal N}_I^{(A)}={\cal N}_I^{(A)}-{\cal N}_0^{(A)}$; $\sum_p {\cal N}^{(p)}_I=N_e$ for all I; In A1N, $N_e=168$ and ${\cal N}_0^{(A)}=92.93$. \gls{DD} refers to visual identification 
from difference density analysis of each state's unrelaxed excited state density at 
density isosurface value of 4$\times$10$^{-4}$ a.u.}\label{tab:a1nlopm}
\begin{tabular}{lcrcrcrc}
\toprule
Functional & $I$ & $\Delta {\cal N}_I^{(A)}$ & & $\omega_I^{(A)} / \omega_I$ & & $\omega_I^{(N)} / \omega_I$ & \gls{DD} \\
\hline
\multirow{5}{*}{LSDA} & 1 & 0.91 & & -76.66~~ && 77.66 ~~& CT: (N)$\rightarrow$(A) \\
                                     & 2 &-0.01 & &   2.27 ~ && -1.27 ~~&  Local on (A) \\
                                     & 3 &-0.90 & &  69.18 ~~&&-68.18 ~~& CT:  (A)$\rightarrow$(N) \\
                                     & 4 & 0.94 & & -59.16 ~~&& 60.16~~ &  CT: (N)$\rightarrow$(A) \\
                                    & 5 &-0.03 & &   2.77 ~~&& -1.77~~ &   Local on (A)\\
\hline
\multirow{5}{*}{PBE1PBE} & 1 & 0.01 & &  1.07~~ & &-0.07~~ & Local on (A)  \\
                                           & 2 & 0.91 & &-56.06~~ && 57.06~~ & CT: (N)$\rightarrow$ (A)  \\
			          & 3 &-0.71 & & 41.90~~ &&-40.90 ~~& CT: (A)$\rightarrow$ (N)  \\
			          & 4 &-0.22 & & 13.22~~ &&-12.22~~ & Delocalized \\
			          & 5 & 0.01 & & -0.54 ~~&&  1.54 ~~& Local on (N)  \\
\hline
\multirow{5}{*}{CAM-B3LYP} & 1 & 0.00 && 1.28 ~~&& -0.28~~ & Local on (A)  \\
                          & 2 &-0.01 && 1.49 ~~&& -0.49~~ & Local on (A)  \\
			   & 3 & 0.71 && -34.75 ~~&& 35.75~~ & CT: (N)$\rightarrow$ (A)  \\
			   & 4 & 0.09 && -4.10 ~~&& 5.10~~  & Delocalized$^{a}$ \\
			   & 5 & 0.05 && -2.19 ~~&& 3.19~~  & Delocalized$^{a}$ \\
\hline
\multirow{5}{*}{$\omega$B97X-D} & 1 & 0.00 & &1.32 ~~&& -0.32~~ &  Local on (A)\\
                                          & 2 &-0.01 & &1.51 ~~&& -0.51 ~~&  Local on (A)\\
			          & 3 & 0.09 &&-4.65~~ &&  5.65 ~~&  Delocalized$^{a}$\\
			         & 4 & 0.01 &&-0.59 ~~&&  1.59 ~~&  Local on (N)\\
			         & 5 & 0.72 &&-34.54~~ && 35.54~~ & CT: (N)$\rightarrow$(A)\\
\hline
\multirow{5}{*}{HF}  & 1 & 0.00 && 1.43~~ && -0.43 ~~& Local on (A)\\
                                   & 2 &-0.01 && 1.51~~ && -0.51~~ & Local on (A)\\
                                   & 3 & 0.00 && 0.01~~ &&  0.99 ~~& Local on (N)\\
                                  & 4 & 0.00 &&-0.15~~ &&  1.15 ~~& Local on (N)\\
                                 & 5 & 0.00 && 1.10 ~~&& -0.10 ~~& Local on (A)\\
\hline
\end{tabular}
\stepcounter{footnote}\footnotetext{The majority of the difference density is on the N-fragment.}
\end{table}
\begin{figure}
\centering
\caption{Orbital energy diagram illustrating occurrence of negative energy differences between partitioned virtual $\epsilon_a$ and occupied $\epsilon_i$ energies: (left) occupied $\phi_i$ and virtual $\phi_a$ orbitals
are localized on donor and acceptor fragments respectively; (right) occupied $\phi_i$ orbital is localized on the B-fragment
and virtual $\phi_a$ orbital is delocalized.}
\includegraphics[width=0.5\textwidth]{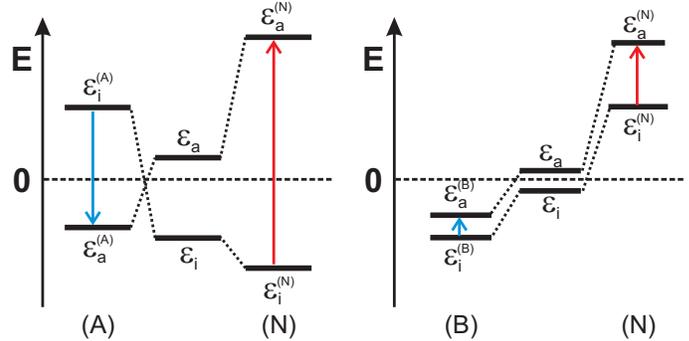}
\label{fig:orbanal}
\end{figure}

\textit{Partitioned excitation energies of A1N:} LOPM characterization of A1N excited state energies 
in the form of ratios between partitioned and total energy differences, $\omega_I^{(p)}/\omega_I$, 
is presented in Table~\ref{tab:a1nlopm}. The magnitude of these ratios reflects the extent of 
excitation energy localization on the fragment. Table~\ref{tab:a1nlopm}
also provides more common analysis of excited states based on the difference between 
unrelaxed one-electron excited densities and the ground state density, we will refer to this approach 
as the difference density (DD) analysis. Combined with an evaluation of the
charge difference on a fragment for the ground and excited states\cite{note2} 
DD is usually used to characterize excitation localization.\cite{wiberg1992,*gordon1995,*nitta2012}  
For localized excitations DD analysis usually predicts changes in density on the fragment where
partitioned energy is localized according to the LOPM. However, there could be exceptions 
to this simple observation as we will see below.   

As discussed in detail in Ref.~\onlinecite{nagesh2015}, 
the \gls{LOPM} approach can produce negative ratios due to de-excitations on a 
fragment $\omega_I^{(p)}<0$. Even from a variational point of view, 
the result $\omega_I^{(p)}<0$ is not surprising
because partitioned energies $E_I^{(p)}$ are expectation values 
of the partitioned Hamiltonian and thus do not have to be ordered in the same way
as the corresponding energies $E_I$ obtained as per the variational principle applied to the full system. 
The largest de-excitations 
are observed in \gls{CT} states; to understand their origin it suffices 
to consider their main components from the orbital energy difference part 
$\Delta E_I\approx\EPS{a}-\EPS{i}$, where orbitals $\phi_a$ and $\phi_i$ are localized on different chromophores.
Following the example given in Fig.~\ref{fig:orbanal}(left) where $\phi_{i}$ is localized on the naphthalene (N) fragment
it is easy to predict that the one-electron part of $\EPS{i}$ upon partitioning will be 
almost completely belong to the N-fragment because of the orbital density localization. 
On the other hand, the N-part of the Coulomb contribution for this orbital energy  
\begin{align}
J^{(N)}(\rho_i,\rho_0) &= \int d\rr d\rr' \frac{\rho_i(\rr')\rho_0(\rr)(\theta_N(\rr')+\theta_N(\rr))}{|\rr'-\rr|} \\
&=\frac{1}{2}[J(\rho_i^{(N)},\rho_0)+J(\rho_i,\rho_0^{(N)})] \label{eq:jsymm}
\end{align}
will be significantly reduced due to the averaging of a large $J(\rho_i^{(N)},\rho_0)$ 
and small $J(\rho_i,\rho_0^{(N)})$ components.
Here, $\rho_i$ and $\rho_i^{(N)}$ are unpartitioned and partitioned orbital densities, and 
$\rho_0$ and $\rho_0^{(N)}$ are corresponding total ground state densities. The reduction 
of $J(\rho_i,\rho_0^{(N)})$ compared to $J(\rho_i^{(N)},\rho_0)$ and $J(\rho_i,\rho_0)$ takes place 
because the total density is always delocalized and its partitioned counterpart is significantly smaller.
This reduction of the Coulomb component in $\EPS{i}^{(N)}$ leads to a decrease of $\EPS{i}^{(N)}$ 
compare to $\EPS{i}$.  Due to the additivity of \gls{LOPM} scheme, $\EPS{i}^{(A)}$ experiences an increase. 
The opposite trend is observed for $\EPS{a}$ due to a different localization of orbital $\phi_a$. 
Overall this produces a large excitation and de-excitation on the N- and A-fragments 
respectively [see Fig.~\ref{fig:orbanal}(left)].

Interestingly, for CT excitations, the fragment losing electronic charge becomes excited. 
This is related to a destabilization of the electron donating fragment since the electron departs from the 
occupied orbital that is below the Fermi level. 

\begin{table}
\caption{Local \gls{TD-DFT} populations and excitation energies in Closs-M using the $\omega$B97X-D functional: $\Delta {\cal N}_I^{(B)}={\cal N}_I^{(B)}-{\cal N}_0$; $\sum_p {\cal N}^{(p)}_I=N_e$ for all I; In Closs-M, $N_e=130$ and ${\cal N}_0^{(B)}=55.00$. \gls{DD} refers to the visual result from difference density analysis of each state's unrelaxed excited state density at density isosurface value of 4$\times$10$^{-4}$ a.u.}\label{tab:trip}
\begin{tabular}{lccrrrc}
\toprule
$I$ & $\omega_I(eV)$ & Spin & {\hspace{0em}}$\Delta {\cal N}_I^{(B)}$ & $\omega_I^{(B)} / \omega_I$ & $\omega_I^{(N)} / \omega_I$ & \gls{DD} \\
\hline
1 & 2.80 & T & 0.00~~ & 0.13~~ & 0.87 ~~& Local on (N) \\
2 & 3.30 & T & 0.00~~ & 1.92 ~~& -0.92~~ &  Local on (B)\\
3 & 3.37 & T &-0.01~~ &-1.18 ~~&  2.18 ~~&  Local on (B)\\
4 & 3.93 & S &-0.01~~ &-1.02 ~~&  2.02~~ &  Local on (B)\\
5 & 4.69 & T & 0.00~~ & 0.01~~ &  0.99~~ &  Local on (N)\\
\hline
\end{tabular}
\end{table}

\textit{Excited states in Closs-M:} To avoid spurious low-energy CT states 
we have used the $\omega$B97X-D functional for this system. 
Table~\ref{tab:trip} shows the partitioned excitation energies 
of low-lying triplet and singlet states in the Closs-M molecule. All excitation energies are 
well-localized, but results of the \gls{DD} analysis disagree with those of the \gls{LOPM} for states 3 and 4. 
These states have similar distributions of single-electron excitation/de-excitation coefficients 
and only differ by the spin multiplicity. 
An examination of the dominant coefficients 
reveals that the occupied \glspl{MO} are localized on the benzaldehyde (B) fragment while 
the virtual \glspl{MO} are delocalized over the whole molecule. 
This 
leads to larger partitioned Coulomb repulsion in $\epsilon_a^{(N)}$ than in $\epsilon_i^{(N)}$
that can be seen from \eq{eq:jsymm} and its counterpart for $\rho_a$:   
$J(\rho_i,\rho_0^{(N)})< J(\rho_a,\rho_0^{(N)})$ since 
$\rho_a$ is partially localized on the N-fragment, hence it interacts stronger with $\rho_0^{(N)}$
than $\rho_i$ does; and  $J(\rho_i^{(N)},\rho_N)< J(\rho_a^{(N)},\rho_0)$ because $\rho_i^{(N)}$ is tiny 
owing to $\rho_i$ localization on the B-fragment. 
Therefore this greater partitioned Coulomb repulsion makes $\epsilon_a^{(N)} - \epsilon_i^{(N)}$ 
positive [see Fig.~\ref{fig:orbanal}(right)] and determines excitation on the naphthalene. 

\section{Concluding remarks} \label{sec:conclusions}
We have developed and implemented a direct scheme of the \gls{LOPM} to partition
the electronic energy of a molecule within the linear 
response \gls{TD-DFT} framework using a combination of numerical and analytical integrals involving a \gls{RI} technique in the \gls{KS} formalism. We apply the \gls{LOPM} to the singlet 
and triplet excited state energies of bichromophore molecules, and find that \gls{LOPM} is a 
powerful method that not only partitions electronic energies in an fragment-additive manner, but also 
provides insights in various one- and two-electron contributions 
of the fragment energies. It was found that a regular density difference approach to excitation analysis 
can be qualitatively misleading for the energy partitioning 
and the \gls{LOPM} provides a quantitative and reliable alternative. 

Further, computed partitioning results do sharply emphasize the well-known need 
for improved functionals, particularly those that effectively treat charge-transfer excitations, 
since there is little consistency amongst partitioning results as the functionals are changed.
However, since the \gls{LOPM} methodology developed here is independent of the functionals adopted, 
the partitioning results are expected to improve as DFT accuracy improves.

\section*{Acknowledgements}
We acknowledge  Mr. Yaser Khan and Professor Moshe Shapiro for stimulating  
discussions at the beginning of this project. The authors also thank V.N. Staroverov 
and I.G. Ryabinkin for helpful discussions. J.N. is very grateful for the hospitality of Gaussian Inc.
A.F.I. greatly appreciates financial support by the Alfred P. Sloan Foundation and 
the Natural Sciences and Engineering Research Council of Canada (NSERC) 
through the Discovery Grants Program. 
P.B.  acknowledges financial support from the Air Force Office of Scientific  Research under Contract No. FA955-13-1-0005.


\begin{thebibliography}{47}
\bibitem [{aun()}]{aunote}%
  \BibitemOpen
  \href@noop {} {}\bibinfo {note} {{A}tomic units are used throughout in this
  work.}\BibitemShut {Stop}%
\bibitem [{not()}]{note2}%
  \BibitemOpen
  \href@noop {} {}\bibinfo {note} {Although in this work we used LOPM
  calculations for charge differences, we do not expect qualitative changes if
  other forms of the density partitioning would be used, for example, Mulliken
  or L\"{o}wdin.}\BibitemShut {Stop}%
\end{thebibliography}
%

\end{document}